\documentclass[a4paper,10pt]{article}

\title{Superloop Space}
\author{Mir Faizal \\ Mathematical Institute, University of Oxford, 
 \\ Oxford,
OX1 3LB, United Kingdom.
}
\date{}
\begin{document}

\maketitle

\begin{abstract}
In this paper will construct and analyse the superloop space formulation of a 
$\mathcal{N} =1$ supergauge theory in three dimensions. 
We will obtain expressions for the connection and the curvature in this superloop space in terms of 
ordinary supergauge fields. This curvature will vanish, unless there is a monopole in the spacetime. 
We will also construct a quantity which will give the monopole charge in this formalism. 
Finally, we will show how these results even hold for a deformed superspace. 
\end{abstract}
\section{Introduction}
In this paper we will construct a superloop space formulation of a $\mathcal{N} =1$ supergauge theory.
To do that we will generalize Polyakov variables to superspace. 
Polyakov variables have been used to construct a loop space formulation of ordinary gauge theories \cite{p1}.
In this formulation, the Polyakov variables take values
in the  Lie algebra and depend on the parametrized loop in the abstract infinite dimensional loop space. 
The loops are formed by first parameterizing the full loop space and then constructing functionals on this parameterized loop space. 
The Polyakov variables act as connection  on loop space and are constructed 
 in analogy with the connections in the  regular gauge theory. Thus, the  Polyakov variables measures the 
 the change in phase  as one moves from one point in the loop space 
to a neighboring point. They can also be used to construct a curvature for the loop space \cite{p2,p3}. This curvature vanishes when the Bianchi 
identities are satisfied. Thus, it vanishes when no monopoles exist. However, when monopoles exist this curvature does not vanish. 
A loop can also be constructed in the loop space \cite{p4}. This loop in the 
loop space   covers a surface in spacetime and can thus be used to 
obtain  the non-abelian monopole charge. These results are know to hold for ordinary gauge theories. 
In this paper we generalize 
these results to three dimensional supergauge theories with $\mathcal{N} =1$ supersymmetry.  
In order to do that we will first review  a superspace formalization of supergauge theories in three dimensions. 

 Supersymmetric gauge theories in three dimensions have 
been throughly studied in $\mathcal{N} =1$ superspace formalism \cite{s,s1,s2, s4}. These theories have become very important due to the discovery of the
BLG  theory  \cite{blg1, blg2, blg3, blg4} and the 
ABJM theory \cite{ab1, ab2, ab3, h1}. These theories are Chern-Simons matter theories which that are thought to be the low energy 
theories for multiple M2 branes. The ABJM theory has only $\mathcal{N} =6$ supersymmetry, however, by the use of the monopole operators
it is possible to obtain the full  $\mathcal{N} =8$ supersymmetry \cite{n1, n2, n3, n4}. 
Thus, starting from  supersymmetric Chern-Simons matter theory
 which is a truncated version of the ABJM theory,
 monopole operator can be used 
to show that  there is an additional ${\cal N}=2$ supersymmetry associated with a particular gauge group.
This additional supersymmetry can combine with ${\cal N}=6$ supersymmetry of the original ABJM theory to an enhanced 
${\cal N}=8$ supersymmetry  with that particular gauge group. 
Hence, it is important to study to effect of monopoles in three dimensions. 

Superspace coordinates for $\mathcal{N} =1$ supersymmetry in three dimensions 
are parameterized by the three dimensional bosonic coordinates $x^\mu$ and a two component 
 Grassman number $\theta^a$.   The spinor indices are raised and
 lowered by $C^{ab}$ and $C_{ab}$, respectively. It is useful to define 
 $x^{ab} = (\gamma^\mu x_\mu)^{ab}\, \partial_{ab} = (\gamma^\mu \partial_\mu)_{ab}$ and $\theta^2 = C^{ab}\theta_a \theta_b/2$. The 
$\mathcal{N} =1$ supersymmetry  in three dimensions is generated by $Q_a = \partial_a - \theta^b \partial_{ab}$. 
This generator of supersymmetry commutes with $D_a = \partial_a + \theta^b \partial_{ab}$.
We now start with a scalar superfield which transforms like 
$
 \delta\psi = i \Lambda \psi,  
$
where $\psi = \psi^A T_A, $ and $[T_A, T_B] = i f_{AB}^C T_C$. Now  we can define a covariant derivative 
for these scalar superfield by requiring the covariant derivative to transforms as the original scalar superfield. 
This covariant derivative is given by 
$\nabla_a = D_a - i \Gamma_a$, where $\Gamma_a = \Gamma_a^A T_A$ is a spinor superfield which transforms   
as $\delta \Gamma_a =  \nabla_a \Lambda$. We also define a vector covariant derivative $\nabla_{ab} = \partial_{ab} - i \Gamma_{ab}$, 
where $\delta \Gamma_{ab} = \nabla_{ab} \Lambda$. We have 
$
 \{\nabla_a, \nabla_b\} = 2i \nabla_{ab},
$
and so $\Gamma_{ab } = - i [D_{(a} \Gamma_{b)} - i \{\Gamma_a, \Gamma_b \}]/2$.
It may be noted that $[\Gamma_{ab}]_| = A_{ab} = (A_\mu\gamma^\mu)_{ab}$, where $|$ indicate that 
we set $\theta^a =0$ and $A^\mu$ is the conventional gauge field in three dimensions. 

From a geometric viewpoint as $z^A = (  x^{ab}, \theta^a)$, so, it is natural to regard $\Gamma_a$ and $\Gamma_{ab}$ 
as components of a superform $\Gamma_A = (\Gamma_{ab}, \Gamma_a )$. So,  we define $D_A = (D_a, \partial_{ab})$
and $ \nabla_A = D_A - i \Gamma_A$. Now we have
$
 [ \nabla_A, \nabla_B \} = T^C_{AB}\nabla_C - i F_{AB}
$.
It is useful to let $T^C_{AB}\nabla_C - i F_{AB} = H_{AB}$.
The Bianchi identities are given by 
$
 [\nabla_{[A}, [\nabla_{B}, \nabla_{c)}\}\} =0, 
$
where $[)$ is the graded anti-symmetrization symbol. It is identical to the anti-symmetrization symbol  but with extra factor of 
$(-1)$ for each pair of interchanged fermionic indices. Thus, the Bianchi identities can be written as
$[\nabla_{[A}, H_{BC)}\} =0$. 
It may be noted that for $\Gamma_{ab}$ to be defined from   
 $\Gamma_a$, we have set $F_{ab} =0$ as a constraint. We now define $W_a$ as
$
 [\nabla_a, \nabla_{bc}] = C_{(ab} W_{c)}, 
$
and so we get 
$
 W_a = \frac{1}{2} D^b D_a \Gamma_a - \frac{i}{2} [\Gamma^b, D_b \Gamma_a] - \frac{1}{6} [\Gamma^b, \{\Gamma_b, \Gamma_a\}]. 
$
Now  $\lambda_a = [W_a]_|$ and $ f_{ab} = [D_{a}W_b]_| = [D_{b}W_a]_|$ is the spinor form of the usual field strength 
$
  (F_{\mu\nu}\gamma^\mu \gamma^\nu)^{ab}_{cd}  
=  \frac{1}{2} \delta^{(a}_{(c} f ^{b)}_{d)}. 
$
It may be noted that 
$
 [\nabla_{cd},\nabla^{ab} ] = -\frac{i}{2} \delta^{(a}_{(c} f^{b)}_{d)}
$.

Different deformations of the superspace occurs due to various backgrounds in the string theory. 
The presence of a constant $NS-NS$  background gives rise to noncommutativity 
 \cite{a, b, c, Chu:1998qz,Chu:1999gi, d} and a $RR$ background gives rise to non-anticommutativity
\cite{Ferrara:2000mm, Klemm:2001yu, e,  g}.
Also, a graviphoton background give rise to a noncommutativity between
spacetime and superspace coordinates \cite{ 1a, 2a, 3a, 4a}.
Noncommutative deformations generated by the $NS-NS$ and graviphoton backgrounds do not break any 
supersymmetry.
As we are studding  $\mathcal{N} =1$ supersymmetric theories in three 
dimensions, any non-anticommutative deformation will break all the supersymmetry. 
So, we will only analyse noncommutative deformation of the superspace. 
Thus, in this paper  will only analyse  a noncommutative deformation of
 superloops formalization of supergauge theories.

\section{Superloop Variables}
In this section we will construct superloop  variables for a three dimensional super-Yang-Mills theory with   $\mathcal{N} =1$  supersymmetry. 
The Lagrangian for this theory is formed  from a combination of the gauge fields and the fermionic fields. 
These fields transforms under the action of the generator $Q_a = \partial_\mu - \theta^b \partial_{ab}$ and these generators 
satisfy the $\mathcal{N} =1$ superalgebra, $\{Q_a, Q_b\} = 2\partial_{ab}$. 
The Lagrangian  for this super-Yang-Mills theory is given by $\mathcal{L} = D^2 [W^2]_|$. 
In component form this Lagrangian is given by $\mathcal{L} = i\lambda^a D^b_a \lambda_b 
- f^{ab} f_{ab}/2$, where $D^b_a = \partial^b_a -i A^b_a $. 
This Lagrangian can be used to calculate the propagators along with  the  Feynman's rules for this theory. Thus, it  can be used 
to obtain  the $S$-matrix for different  physical processes. 
However, there are some interesting physical phenomena like the Aharonov-Bohm effect \cite{abaa, abaa1, abaa2, abaa4} where 
the global properties of spatial regions 
can effect physics in local gauge theories. Thus, in order to capture such effect in super-Yang-Mills theories we need to construct 
a superloop formalization for them.  

 In 
order to do that we first construct the coordinates for each point in the superloop space.  
These coordinates parameterizing the superloop space are $\xi^A 
= (\xi^{ab}, \xi^a)$, 
\begin{equation}
 C : \{ \xi^A (s): s = 0 \to 2\pi, \, \, \xi^A (0) = \xi^A(2\pi)\},  
\end{equation}
where   $\xi^A (0) = \xi^A(2\pi)$ is a fixed point in the superloop space. Now we can define a superloop 
variable as functional on the set of all such functions 
\begin{equation}
 \Phi [\xi] 
= P_s \exp i \int^{2\pi}_0  \Gamma^A (\xi(s)) \frac{d \xi_A}{ds}, 
\end{equation}
where 
\begin{equation}
 \Gamma^A (\xi(s))  \frac{d \xi_A}{ds} =
 \Gamma^{ab} (\xi(s))  \frac{d \xi_{ab}}{ds} +  \Gamma^a (\xi(s))  
\frac{d \xi_a}{ds},
\end{equation}
  and $P_s$ denotes ordering in $s$ 
increasing from right to left. The derivative in $s$ is  taken from below.
It may be noted that 
$\Phi[\xi] $ is a scalar superfield from the supersymmetric point of view, 
\begin{eqnarray}
[\Phi[\xi]]_| = \phi[\xi], && [D_a\Phi[\xi]]_| = \phi_a[\xi], \nonumber \\ 
{[D^2\Phi[\xi]]_| }= \tilde\phi[\xi],&& \label{a}
\end{eqnarray}
where $\phi[\xi], \, \phi_a[\xi], \, \tilde\phi[\xi] $ are loop variables formed from the component 
fields of the super-Yang-Mills theory.  
 They are thus regular loops formed from 
linear combinations of various field that exists in the super-Yang-Mills theory.  
Now using $\Phi[\xi]$, we can define 
\begin{equation}
 F_A [\xi| s] = i \Phi^{-1}[\xi] \frac{\delta}{\delta \xi^A (s)}\Phi[\xi]. 
\end{equation}
 Here each of these components of 
 $ F_A [\xi| s] = (F_{ab}[\xi| s],
 F_a[\xi| s])$ is  obtained by taking  a vector or a spinor derivative.
This equation can be understood as a parallel phase transport first  to $\xi[s]$ along some path, followed by 
a detour at $s$, and then backward along the same path. The phase factors generated by first going  forward and then 
going backward from $s$ cancel each other. However, the phase factor generated by taking a detour at $s$ generates 
$H^{AB}$ because of the transport along the infinitesimal circuit at $s$. Thus, we can write 
\begin{equation}
 F^A [\xi|s] =   \Phi^{-1}(\xi: s,0) H^{AB} (\xi (s) )\Phi^{-1}(\xi: s,0)\frac{d \xi_B (s) }{d s}, \label{la}
\end{equation}
 where 
 \begin{equation}
 \Phi [\xi: s_1, s_2 ] 
= P_s \exp i \int^{s_2}_{s_1}  \Gamma^A (\xi(s)) \frac{d \xi_A}{ds},
\end{equation}
is the parallel transport from a point $\xi(s_1)$ to $\xi(s_2)$ along path parametrized by $\xi$. 
Thus,  we parallel transport first forward to $s$ and then take a detour at $s$ and then turn backwards again 
along the same path. The phase factor for the segment of the superloop beyond $s$ cancels and 
the factor for the remainder do not. The detour at $s$ gives a truncated phase factors and the infinitesimal circuit 
generated at $s$ gives rise to $H^{AB}(\xi (s) )$. 
It may be noted that $ F^A [\xi|s]$ is a connection in the superloop space and not in spacetime. 
It is propotional to the field strength in spacetime. In the next section we will show that 
this connection is flat as the field strength 
corresponding to it vanishes due to the Bianchi indenity. However,  in presence of a monopole 
Bianchi indentity do not hold, so this connection is not flat in presence of a monopole. 

Now as $ F^A [\xi|s] $ represents the change in phase of $\Phi[\xi]$ as one moves from one point in the superloop space 
to a neighboring point, we can regard it as a connection in the superloop space.  Thus, in analogy with 
ordinary gauge theories, we can proceed to construct a curvature for the superloop space. 
The local change in the phase as a point moves around an infinitesimal 
closed circuit in superloop space will now be given by the curvature $G_{AB} [\xi, s_1, s_2]$, where 
\begin{eqnarray}
 G_{AB}[\xi, s_1, s_2] &=& \frac{\delta}{\delta \xi^B (s_2) }F_A [\xi|s_1]
- \frac{\delta}{\delta \xi^A (s_1) }F_B [\xi|s_2] \nonumber \\&&
+i [F_A [\xi|s_1], F_B [\xi|s_2]].
\end{eqnarray}

\section{Monopoles}
In this section we will show that 
the loop space curvature vanishes unless monopoles are present in the spacetime. 
To obtain this result,  we first have to express the connection 
in superloop space in terms of  usual field variable. In order to do that we 
first evaluate the value of $ \Phi^{-1}[\xi_2] \Phi[\xi_3] - \Phi^{-1}[\xi] \Phi[\xi_1]$, where 
$
  \xi_3^A (s) = \xi_1^A (s) + \delta \xi'^A(s),\, \, 
  \xi_2^A (s) = \xi^A (s) + \delta \xi'^A(s), $ and $ \xi_1^A (s) = \xi^A (s) + \delta \xi^A(s) $.
By repeating the argument used in the derivation of Eq. (\ref{la}), we obtain the following result
\begin{eqnarray}
 \Phi[\xi_1] = \Phi[\xi] - i \int ds \Phi( \xi: 2\pi, s ) H^{AB} (\xi(s)) \frac{d \xi_B(s)}{ds} \nonumber \\ 
\times \delta \xi_A (s) \Phi(\xi: s, 0),
\end{eqnarray}
A similar  expression can be obtained for $\Phi[\xi_2]$. Furthermore, we also have 
\begin{eqnarray}
  \Phi[\xi_3] = \Phi[\xi_1] - i \int ds \Phi( \xi_1: 2\pi, s ) H^{AB} (\xi_1(s)) \frac{d \xi_{1B}(s)}{ds} \nonumber \\ 
\times \delta \xi_{1A} (s) \Phi(\xi_1: s, 0),
\end{eqnarray}
Now can write 
\begin{eqnarray}
\Phi(\xi_1: s, 0) = \Phi(\xi: s, 0) - i \int_0^s ds' \Phi( \xi: s, s' ) H^{AB} (\xi(s')) \frac{d \xi_B(s')}{ds'} \nonumber \\ 
\times \delta \xi_A (s') \Phi(\xi: s', 0) + i \Gamma^{A}(\xi(s)) \Phi(\xi: s, 0)\delta \xi_A (s).
\end{eqnarray}
Here the last term is due to the variation of the end-point in the integral for $\Phi(\xi: s,0)$. 
A similar expression can be written for $\Phi(\xi: 2\pi, s)$.
Collecting all the variations, we can write  
\begin{eqnarray}
 \frac{\delta}{\delta \xi_B (s_2) }F^A [\xi|s_1] &=& i [F^B [\xi|s_2], F^A [\xi|s_1]] \delta (s_1-s_2) \nonumber \\ 
&&+  \Phi^{-1}( \xi: s_1, 0 ) \nabla^A H^{BC} (\xi(s)) \Phi(\xi: s_1, 0)\nonumber \\ 
&&\times \frac{d \xi_B(s_1)}{ds_1}\delta \xi_{1C} (s_1)\delta (s_1-s_2) \nonumber \\ 
&&+ \Phi^{-1}( \xi: s_2, 0 )  H^{BC} (\xi(s_2)) \Phi(\xi: s_2, 0)\nonumber \\ 
&&\times\frac{d }{ds_2}\delta \xi_{1C} (s_1)\delta (s_1-s_2).
\end{eqnarray}
Thus, we get
\begin{eqnarray}
 G_{AB}[\xi, s_1, s_2] &=& \Phi^{-1}(\xi: s_1,0) [\nabla_{[A}, H_{BC)}\}
  \nonumber \\ && \Phi(\xi: s_1,0)\frac{d\xi^C (s_1)}{ds_1}\delta (s_1-s_2). 
\end{eqnarray}
Now if the  Banichi identity hold, $[\nabla_{[A}, H_{BC)}\} =0$, then, $ G_{AB}[\xi, s_1, s_2] = 0$. 
It may also be noted that this curvature is proportional to $\delta(s_1-s_2)$. 

We have now seen  that the curvature of the superloop space vanishes if the Bianchi 
identity holds. However, if a monopoles exists then at places where the superloop space intersects with the world-line 
of a monopole, the Bianchi identity need not hold. Thus, if a monopoles exists the curvature tensor of
the superloop space will not vanish. So, if monopoles are present then $[\nabla_{[A}, H_{BC)}\} \neq 0$, and thus,  
$ G_{AB}[\xi, s_1, s_2] \neq 0$. In other words if $ G_{AB}[\xi, s_1, s_2] \neq 0$ then  the superloop  is 
intersecting word-lines of a monopole.

In order to analyse this further, we define a loop in the superloop space, as 
follows, 
\begin{equation}
  \Sigma : \{ \xi^A (t:s), \, s = 0 \to 2 \pi, \, t = 0 \to 2 \pi\},
\end{equation}
where    
\begin{eqnarray}
\xi^A (t:0) =  \xi^A (t:2\pi), && t  = 0 \to 2 \pi, \nonumber \\ 
\xi^A (0:s) =  \xi^A (2\pi:s), && s  = 0 \to 2 \pi. 
\end{eqnarray}
Thus, for each $t$,  $\xi^A (t:s)$ represents a closed superloop passing through a fixed point. 
As $t$ varies a curve in the superloop space is constructed. Now we can define a loop variable for this space as, 
\begin{equation}
\Theta (\Sigma) = P_t \exp  i \int^{2\pi}_0 dt \int^{2\pi}_0 ds
F^A (\xi(t: s)) \frac{\partial \xi_A (s)}{ \partial t}, 
\end{equation}
and $P_t$ denotes ordering in $t$ increasing from right to left and the derivative is taken from below. 
It may be noted that 
$\Theta (\Sigma) $ is also a scalar superfield from the supersymmetric point of view, 
\begin{eqnarray}
[\Theta (\Sigma)]]_| = \theta (\Sigma), && [D_a\Theta (\Sigma)]_| = \theta_a (\Sigma), \nonumber \\ 
{[D^2\Theta (\Sigma)]_| }= \tilde\theta (\Sigma),&& 
\end{eqnarray}
where $\theta (\Sigma), \, \theta_a (\Sigma), \, \tilde\theta (\Sigma) $ are loop variables formed from the component 
fields of the super-Yang-Mills theory. 

Here $F^A[\xi|s]$ plays the role of connection, with the difference that it is infinite dimensional. Thus, apart from 
the sum over $\mu$, we have to also integrate over $s$. 
In spacetime this loop in superloop space is generated by a parametrized two dimensional surface, enclosing a three dimensional 
volume. Now as $F^A[\xi|s]$ is generated by derivative of $\Phi[\xi]$, so $\Theta (\Sigma)$ measures total change in $\Phi[\xi:t]$ as 
$t$ varies from $t = 0 \to 2\pi$. Now $\Theta (\Sigma)$ is again an element of gauge group, say $SU(2)$. 
If $\Sigma$ encloses a monopole, then $t  = 0 \to 2 \pi$ will only trance curve in $SU(2)$ which winds only half way around the group. 
 If $\Sigma $ does not include a monopole then it traces out a closed curve in $SU(2)$. 
As $\Theta (\Sigma)$ measure the total change,  it is proportional to the monopole charge.
Thus, the monopole charge of a $SU(2)$ theory is $\pm 1$.

\section{Deformed  Superspace }
In this section we shall generalize the results of the previous sections to  a noncommutative  deformation of the superspace. 
In order to analyse deformation 
of the superspace both the Grassman 
coordinates and the spacetime coordinates are
 promoted to operators and a deformation of 
there superalgebra is imposed. 
Thus,  we  promote ${ \theta}^a$ and ${y}^\mu$ 
 to operators $\hat{ \theta}^a$ and $\hat{y}^\mu$ 
which satisfy the following superspace algebra, 
\begin{eqnarray}
  [\hat{y}^\mu, 
\hat{y}^\nu] = 
 B^{\mu\nu}, &&
 {[\hat{y}^\mu, \hat{\theta}^a]}= A^{\mu a }.  
\end{eqnarray} 
This deformation induces  the following star product
  between  functions of ordinary superspace \cite{ab3,h1},
\begin{eqnarray}
{\Gamma^A}(y,\theta) \star { \Gamma_{A }}  (y,\theta) & =& 
\exp -\frac{i}{2} \left(
 B^{\mu\nu} \partial^2_\mu \partial^1_\nu
+ A^{\mu a} (D^2 _a \partial^1_\mu -\partial^2_\mu 
D^1_a \right)) \nonumber \\ &&
\, \,\,\,\,\,\, \,\,\,\,\,\,
 \times 
 {\Gamma^A}(y_1,\theta_1) { \Gamma_{A  }}  (y_2, \theta_2)
\left. \right|_{y_1=y_2=y, \; \theta_1=\theta_2=\theta}.
\label{star2}
\end{eqnarray}
It may be noted that if we deform this algebra by $\{\theta_a, \theta_b \} = C_{ab}$, we will 
break all the supersymmetry of the theory. However, in  four dimensions or for $\mathcal{N} =2$ supersymmetry 
in three dimensions, this deformation can be performed \cite{Ferrara:2000mm, Klemm:2001yu, e,  g}. 
Here we have defined the star product between ordinary functions using 
super-derivative $D_a$ rather than $\partial_a$ because they commute 
with the generators of the supersymmetry $Q_a$. 

In this deformed superspace a deformed 
version of Bianchi identity is satisfied $[\nabla_{[A}, H_{BC\star)}\}_\star =0$, where 
$H_{BC \star} = [ \nabla_A , \nabla_C \}_\star $.
It may be noted that for $\Gamma_{ab}$ to be defined from   
 $\Gamma_a$, we have to again  set $F_{ab} =0$ as a constraint. Now  we have 
$
 [\nabla_a, \nabla_{bc}]_{\star} = C_{(ab} W_{c)}, 
$
and so we get 
$
 W_a = \frac{1}{2} D^b D_a \Gamma_a - \frac{i}{2} [\Gamma^b, D_b \Gamma_a]_{\star} - \frac{1}{6} [\Gamma^b, \{\Gamma_b, \Gamma_a\}_{\star}]_{\star}. 
$
The Lagrangian for the deformed super-Yang-Mills theory is  given by 
$
\mathcal{L}= D^2  [W^a \star W_a]_|.
$
After analysing the deformation of the superspace, we can construct a superloop space formalization 
for the deformed gauge theory on it. Thus, 
 we can define a superloop variable as
\begin{equation}
 \Phi_\star [\xi] 
= P_s \left[ \exp i \int^{2\pi}_0  \Gamma^A (\xi(s))  \frac{d \xi_A}{ds} \right]_\star, 
\end{equation}
where 
\begin{equation}
 C : \{ \xi^A (s): s = 0 \to 2\pi, \, \, \xi^A (0) = \xi^A(2\pi)\}.  
\end{equation}
Here  all the products of fields inside the brackets are taken as star products.
It may be noted that 
$\Phi_\star [\xi] $ is also scalar superfield from the supersymmetric point of view, 
\begin{eqnarray}
[\Phi_\star [\xi]]_| = \phi_\star [\xi], && [D_a\Phi_\star [\xi]]_| = \phi_{a\star}[\xi], \nonumber \\ 
{[D^2\Phi_\star [\xi]]_| }= \tilde\phi_\star [\xi],&& 
\end{eqnarray}
where $\phi_\star [\xi], \, \phi_{a\star}[\xi], \, \tilde\phi_\star [\xi] $ are loop variables formed from a different combination 
of the component 
fields of the super-Yang-Mills theory, as compared to Eq. (\ref{a}). These loop variables exist  on noncommutative 
spacetime. 
We can now define a connection for this deformed superloop space as
\begin{equation}
 F^A_{\star} =   \Phi^{-1}_\star(\xi: s,0) \star H^{AB}_\star (\xi (s) )\star \Phi^{-1}_\star(\xi: s,0) \frac{d \xi_B (s) }{d s} ,
\end{equation}
Here again $ F^A_{\star} [\xi|s] $ represents the change in phase of $\Phi_\star [\xi]$ as one moves from one point in the deformed superloop space 
to a neighboring point in it.  

Now we can again  
 construct a curvature for the deformed superloop space by replacing all the products of fields by star products. 
Thus, the curvature in this deformed superloop space is given by 
\begin{eqnarray}
 G_{AB\star}[\xi, s_1, s_2] &=& \frac{\delta}{\delta \xi^B (s_2) }  F_{A \star} [\xi|s_1] 
- \frac{\delta}{\delta \xi^A (s_1) }F_{B \star} [\xi|s_2] \nonumber \\&&
+i [F_{A \star} [\xi|s_1], F_{B \star} [\xi|s_2]]_\star.
\end{eqnarray}
Repeating the above argument with star-product replacing the ordinary product, we obtain 
\begin{eqnarray}
 G_{AB\star}[\xi, s_1, s_2] &=& \Phi^{-1}_\star(\xi: s_1,0) \star[\nabla_{[A}, H_{BC\star)} \}_\star
  \nonumber \\ && \star \Phi_\star(\xi: s_1,0)\frac{d\xi^C (s_1)}{ds_1}\delta (s_1-s_2).
\end{eqnarray}
If there are no monopoles in the spacetime, then   the deformed  Bianchi identity holds, $[\nabla_{[A}, H_{BC\star) } \}_\star =0$, 
and thus, $ G_{AB\star}[\xi, s_1, s_2] = 0$. To analyse the effect of monopoles we again 
 define 
\begin{equation}
\Theta_{\star} (\Sigma) = P_t \left[ \exp  i \int^{2\pi}_0 dt \int^{2\pi}_0 ds
F^A_{\star} (\xi(t: s)) \frac{\partial \xi_A (s)}{ \partial t} \right]_\star, 
\end{equation}
where    
\begin{eqnarray}
\xi^A (t:0) =  \xi^A (t:2\pi), && t  = 0 \to 2 \pi, \nonumber \\ 
\xi^A (0:s) =  \xi^A (2\pi:s), && s  = 0 \to 2 \pi. 
\end{eqnarray}
We can obtain the component fields   $\theta_{\star} (\Sigma), \, \theta_{a\star} (\Sigma), \, \tilde\theta_{\star} (\Sigma) $
from $\Theta_{\star} (\Sigma)$, just as we obtained the component fields of $\Phi_{\star}[\xi]$. These component fields also exist 
on noncommutative spacetime.  
Now by repeating the above argument  for a monopole in this deformed theory, we can  show that the
monopole charge of a $SU(2)$ theory on this deformed superspace is again $\pm 1$. 
Thus, all the results of ordinary superloop space hold  even after deforming the 
superspace, with the only difference that the ordinary product of fields is converted into the star product. 
\section{Conclusion}
In this paper we first constructed a superloop space formulation of super-Yang-Mills theory. 
This was done by defining a superloop variable that was a superscalar field from the view point of 
supersymmetry. This was formed by a linear combination of the component fields. Then a connection 
on this superloop space was constructed by taking spinor and vector derivatives of this quantity. Finally, 
a curvature on this superloop space was also constructed. This curvature vanished if there was no monopole in the 
spacetime. However, if a monopole existed in the spacetime and the superloop passed through its world lines, then this curvature did not 
vanish. We also constructed a quantity that would measure the monopole charge by constructing loops of superloop space. 
This two dimensional quantity measured the monopole charge. Finally, it was shown that all these results hold even after deforming 
the superspace. 

In abelian gauge theories  a duality exists which is generated by the Hodge star operation. 
This duality cannot be generalized in a straight forward way to non-abelian gauge theories. However, 
in the loop space formulation of Yang-Mills theories, this duality has been generalized 
and  a non-abelian generalized dual transform has been constructed \cite{p5, p6}. 
It will be interesting to generalize these results to super-Yang-Mills theories. This can be done by first using 
the results of this paper and constructing generalized duality transformations in three dimensions. After that it will be interesting 
to generalize the results of this paper to $\mathcal{N} =1$ supersymmetric Yang-Mills theory in four dimensions or 
a $\mathcal{N} =2$ supersymmetric Yang-Mills theory in three dimensions and obtain generalized duality transformations for them. 
Lastly, it will also be interesting to analyse the ABJM theory with monopole operators in this formalism.
It may be noted that so far the analyses of a non-abelian two form gauge field in loop space has not been performed. 
It will be interesting to construct a loop space formalization of this field. It could be possible to do so by using 
the concept of parametrized surfaces.
 It might also then become possible to study a theory of multiple $M5$ branes  using this formalism \cite{m5aa, m51a, m52a, m1a}. 
 
\section*{Acknowledgment}
I am grateful to Tsou Sheung Tsun for useful discussions. 

\end{document}